\documentclass[conference]{IEEEtran}
\usepackage[utf8]{inputenc}
\usepackage{graphicx}
\usepackage{amsmath}
\usepackage{booktabs}
\usepackage{array}
\usepackage{caption}
\usepackage{subcaption}
\usepackage[hyphens]{url}
\usepackage[colorlinks=true,allcolors=blue]{hyperref}

\title{A Comparative Performance Evaluation of Kyber, sntrup761, and FrodoKEM for Post-Quantum Cryptography}
\author{
    \IEEEauthorblockN{Samet Ünsal}
    \IEEEauthorblockA{
        Department of Computer Engineering\\
        Izmir Katip Celebi University\\
        Izmir, Turkiye\\
        Email: 220401082@ogr.ikcu.edu.tr
    }
}

\begin{document}
\maketitle

\begin{abstract}
Post-quantum cryptography (PQC) aims to develop cryptographic algorithms that are secure against attacks from quantum computers. This paper compares the leading post-quantum cryptographic algorithms, such as Kyber, sntrup761, and FrodoKEM, in terms of their security, performance, and real-world applicability. The review highlights the strengths and weaknesses of each algorithm and provides insights into future research directions. We also discuss the challenges of transitioning from classical to post-quantum systems and the potential impacts on various industries. This paper serves as a foundation for understanding the current state of post-quantum cryptography and its future prospects in the quantum computing era.
\end{abstract}

\begin{IEEEkeywords}
Post-Quantum Cryptography, PQC, Kyber, sntrup761, FrodoKEM, Quantum Security, Lattice-based Cryptography, Quantum-resistant Algorithms, Security Algorithms, Cryptography in the Quantum Era.
\end{IEEEkeywords}

\section{Introduction}
Quantum computers have made tremendous progress in recent years, and their potential capabilities could revolutionize many fields of science and technology. However, one of the most significant challenges posed by the advent of quantum computing is its impact on \textbf{cryptography}, which underpins the security of modern communication systems. Traditional cryptographic algorithms, such as \textbf{RSA} and \textbf{Elliptic Curve Cryptography (ECC)}, rely on mathematical problems that are computationally hard to solve for classical computers, such as integer factorization and the discrete logarithm problem. These algorithms, however, are vulnerable to quantum algorithms like \textbf{Shor's algorithm}, which can efficiently solve these problems in polynomial time, rendering the classical cryptographic systems insecure in the quantum era \cite{Shor1994}.

The threat posed by quantum computing to existing cryptographic systems has sparked significant interest in the development of \textbf{Post-Quantum Cryptography (PQC)}, which aims to design cryptographic algorithms that remain secure even in the presence of a quantum adversary. PQC focuses on mathematical problems that are believed to be resistant to quantum attacks, such as \textbf{lattice-based cryptography}, \textbf{code-based cryptography}, and \textbf{multivariate polynomial cryptography}. These new cryptographic frameworks provide an avenue for securing data against future quantum computers and are considered essential to the development of quantum-safe communication systems \cite{NIST2020}.

The need for post-quantum cryptographic algorithms is becoming increasingly urgent as quantum computers continue to advance. Current cryptographic systems that rely on classical hardness assumptions, such as RSA and ECC, are expected to be easily compromised once sufficiently powerful quantum computers are developed. The transition to PQC is not only necessary for ensuring the security of sensitive information but also for safeguarding the future of secure communication in various sectors, including finance, healthcare, government, and military.\cite{NIST2020}

In addition to their security considerations, the practical performance of post-quantum algorithms plays a crucial role in their adoption. These algorithms must be evaluated not only in terms of their resistance to quantum attacks but also in their efficiency, scalability, and real-world applicability. Kyber, sntrup761, and FrodoKEM are some of the most promising post-quantum algorithms that have been developed and analyzed in the context of both security and performance. Understanding the strengths and weaknesses of these algorithms, as well as their suitability for implementation in real-world systems, is vital for their successful integration into the global cryptographic infrastructure.\cite{NIST2020}

This paper aims to provide a comprehensive comparison of leading post-quantum cryptographic algorithms, focusing on their security, performance, and potential applications in various industries. We will examine the existing literature on the subject, highlight the key challenges and opportunities associated with post-quantum cryptography, and discuss the future directions of research in this area. Furthermore, we will explore the implications of transitioning from classical to post-quantum systems, considering both the technical and societal aspects of this transition.\par

As quantum computers continue to evolve, it is essential that cryptographic systems be prepared to withstand the challenges posed by this new technology. This paper serves as a foundation for understanding the current state of post-quantum cryptography and its future prospects, offering insights into how the cryptographic community is addressing one of the most significant technological shifts of our time.

\section{Literature Review}
\subsection{Classical Cryptographic Systems: RSA, ECC, and AES}
Traditional cryptographic algorithms such as \textbf{RSA}, \textbf{Elliptic Curve Cryptography (ECC)}, and \textbf{Advanced Encryption Standard (AES)} are widely used for securing digital communication, banking, and data transmission. These cryptosystems rely on computationally hard problems such as \textbf{integer factorization} for RSA, the \textbf{discrete logarithm problem} for ECC, and \textbf{block ciphers} for AES. However, these algorithms were designed under the assumption that classical computers would be used for encryption and decryption tasks, which has been fundamentally challenged by the advent of \textbf{quantum computers} \cite{Shor1994}.

\subsubsection{RSA}
RSA is a public-key cryptosystem based on the difficulty of factoring large integers. It is widely used for securing online transactions, email encryption, and digital signatures. The security of RSA is based on the assumption that factoring large composite numbers is computationally infeasible. However, quantum algorithms like Shor’s algorithm can factor large numbers in polynomial time, rendering RSA insecure against quantum attacks \cite{Shor1994}. This vulnerability highlights the need for post-quantum cryptographic systems that can resist such quantum attacks.

\subsubsection{Elliptic Curve Cryptography(ECC)}
ECC is a form of public-key cryptography that relies on the algebraic structure of elliptic curves over finite fields. ECC provides the same level of security as RSA but with much smaller key sizes, making it more efficient for use in constrained environments like mobile devices. Like RSA, the security of ECC is based on the discrete logarithm problem, which quantum computers can solve efficiently using Shor’s algorithm \cite{Koblitz1987}. As a result, ECC is also vulnerable to quantum attacks, leading to the need for quantum-resistant alternatives.

\subsubsection{AES and Symmetric Cryptograph}
While RSA and ECC are asymmetric encryption systems, AES is a symmetric encryption algorithm widely used for encrypting data in bulk. AES is generally considered secure against classical attacks, with key sizes of 128, 192, and 256 bits. However, quantum computers can break symmetric encryption through Grover’s algorithm, which reduces the effective security of symmetric ciphers by half. This means
 that a 128-bit key would provide only 64-bit security against a quantum adversary \cite{Grover1996}. As such, post-quantum cryptography also considers the need for stronger symmetric encryption protocols, potentially using larger key sizes to counter quantum attacks.

\subsection{Post-Quantum Cryptographic Algorithms}
As the quantum computing revolution progresses, the classical algorithms mentioned above are likely to become obsolete. To address this, researchers have developed various \textbf{post-quantum cryptographic algorithms} that aim to remain secure in the presence of quantum computers. These algorithms rely on mathematical problems that are believed to be difficult even for quantum computers, such as \textbf{lattice-based cryptography}, \textbf{code-based cryptography}, and \textbf{multivariate polynomial cryptography}.

\subsubsection{Kyber}
Kyber is a lattice-based cryptographic algorithm that has been widely studied due to its security and efficiency. It is a key encapsulation mechanism (KEM) that is resistant to quantum attacks. The algorithm’s security is based on the learning with errors (LWE) problem, which is believed to be hard even for quantum computers \cite{Chen2016}. Kyber has been selected as one of the candidates in the NIST post-quantum cryptography standardization project and is considered a strong contender for future secure communication systems.\cite{NIST2020}

\subsubsection{sntrup761}
sntrup761 is a lattice-based public-key encryption scheme optimized for efficiency in constrained environments. It is part of the NTRU Prime family and has been proposed as an alternative candidate in the NIST Post-Quantum Cryptography standardization project. The security of sntrup761 is based on hard problems in structured lattices, particularly the Ring-LWE and related algebraic lattice problems, which are believed to be resistant to both classical and quantum attacks.\cite{bernstein2018ntruprime}

\subsubsection{FrodoKEM}
FrodoKEM is a post-quantum key exchange protocol that is based on the Learning With Errors (LWE) problem. Unlike Kyber, which relies on structured lattices, FrodoKEM uses unstructured lattices, providing a different approach to achieving post-quantum security. This difference makes FrodoKEM more resistant to certain quantum algorithms and attacks that might affect structured lattice-based schemes.\cite{Booth2018}

\subsection{The Need for Transition to Post-Quantum Systems}
With the rapid progress of quantum computing, the need to transition to post-quantum cryptographic systems has never been more pressing. Traditional cryptosystems like RSA and ECC are expected to be broken by quantum algorithms once sufficiently powerful quantum computers are available. The transition from classical to post-quantum systems involves not only adopting quantum-resistant algorithms but also addressing compatibility issues with existing infrastructures and optimizing performance for real-world applications.
This transition presents a number of challenges:
\begin{itemize}
    \item Security Concerns: Ensuring that post-quantum algorithms provide the same level of security as classical systems.

    \item Key Sizes and Efficiency: Many post-quantum algorithms require significantly larger key sizes and have higher computational costs compared to classical algorithms.

    \item Standardization: The NIST Post-Quantum Cryptography Project has been working on standardizing algorithms that are secure against quantum attacks, providing guidelines for their implementation.\cite{NIST2020}
    
\end{itemize}

\section{Methodology}
In this section, we describe the methodology employed to evaluate the security and performance of leading post-quantum cryptographic algorithms: \textbf{Kyber}, \textbf{sntrup761}, and \textbf{FrodoKEM}. The comparison focuses on key metrics such as \textbf{encryption/decryption time}, \textbf{memory consumption}, and \textbf{security} under quantum attacks. We also discuss the environment used for running these evaluations.

\subsection{Algorithm Selection}
The algorithms selected for evaluation---\textbf{Kyber}, \textbf{sntrup761}, and \textbf{FrodoKEM}---represent a diverse set of approaches to post-quantum cryptography. Each algorithm is based on different mathematical problems that are resistant to quantum attacks. Kyber and sntrup761 are \textbf{lattice-based}, while FrodoKEM uses the \textbf{Learning With Errors (LWE)} problem. These algorithms were selected based on their inclusion in the \textbf{NIST Post-Quantum Cryptography (PQC) Standardization Process}, their demonstrated security, and their applicability to real-world systems \cite{NIST2020}.

\subsection{Test Environment}
The algorithms were implemented using \textbf{Python} with the \textbf{PyCryptodome} library for classical encryption functionalities and custom implementations for post-quantum cryptographic algorithms. The benchmarking was performed on a \textbf{Windows 11 machine} with the following specifications:

\begin{itemize}
    \item \textbf{Processor}: Intel Core i5-13th Gen (12 cores, 16 threads, 3.0 GHz base clock, turbo up to 4.6 GHz)
    \item \textbf{Memory}: 16 GB DDR4 (3200 MHz)
    \item \textbf{Storage}: 512 GB NVMe SSD
    \item \textbf{Python Version}: 3.9.7
\end{itemize}

The scipy library was used for performance analysis, including timing the key generation, encryption, and decryption processes

\subsection{Data Sets and Simulations}
Since the focus of this research is on post-quantum cryptography algorithms, no direct real-world data set is required. Instead, the algorithms were tested using standard cryptographic operations, which include:
\begin{itemize}
    \item \textbf{Key Generation:} Generation of public and private key pairs.
    \item \textbf{Encryption:} Encrypting randomly generated 512-bit messages.
    \item \textbf{Decryption:} Decrypting the encrypted messages and comparing the result with the original plaintext.
\end{itemize}
\par
We performed 400 encryption and decryption operations for each algorithm, measuring the average time per operation, memory usage, and the number of cycles it took for successful encryption and decryption.

\subsection{Security Analysis}
To assess the security of the post-quantum cryptographic algorithms, we simulate the brute-force attack scenario. The security is evaluated by examining 
the resilience of these algorithms against quantum adversaries, specifically focusing on how they would perform against Shor’s algorithm and Grover’s algorithm.
\par
For each algorithm, we tested the encryption and decryption time under a brute-force attack and compared how long it would take to break the system using classical and quantum techniques. We simulated both a classical brute-force attack and a quantum-based attack (using Grover’s algorithm for symmetric ciphers and Shor’s algorithm for public-key encryption).

\subsection{Performance Metrics}
The algorithms were evaluated using the following performance metrics:
\begin{itemize}
    \item \textbf{Key Generation Time:} The time taken to generate the public and private key pairs.

    \item \textbf{Encryption Time:} The time taken to encrypt a 512-bit message.

    \item \textbf{Decryption Time:} The time taken to decrypt the encrypted message
    
\end{itemize}
\par
These metrics are essential for understanding not only the security of each algorithm but also their feasibility for real-world applications.

\subsection{Comparative Analysis}
The performance and security of Kyber, sntrup761, and FrodoKEM are compared based on the following criteria:
\begin{itemize}
    \item \textbf{Security:} The ability to resist known quantum algorithms, including Shor’s algorithm for 

    \item \textbf{Efficiency:} The time and memory required to generate keys, encrypt, and decrypt data.

    \item \textbf{Scalability:} How the algorithms perform when scaling up the message sizes and key sizes.
\end{itemize}
\par
The performance tests include multiple rounds of encryption and decryption with different randomly generated messages, and the security tests focus on ensuring that no algorithm can be broken by classical or quantum brute-force attacks.

\subsection{Challenges and Limitations}
While these algorithms have been selected and tested based on theoretical and practical considerations, there are some challenges:
\begin{itemize}
    \item \textbf{Key Size and Efficiency:} Post-quantum algorithms tend to have larger key sizes and require more computational resources than classical algorithms like RSA and ECC.
    
    \item \textbf{Scalability:} As the size of the messages and keys increases, performance may degrade, posing challenges for large-scale implementations.

    \item \textbf{Quantum Vulnerabilities:} While these algorithms are considered quantum-resistant, future advancements in quantum computing could expose vulnerabilities not yet discovered.
    
\end{itemize}

\subsection{Future Directions}
Future research will focus on improving the efficiency and scalability of these post-quantum cryptographic algorithms, particularly in constrained environments. The impact of quantum attacks on these algorithms will continue to be a major research area, as more powerful quantum computers emerge. Additionally, continued efforts in standardization and global adoption of post-quantum cryptographic standards are necessary for building secure quantum-resistant systems.

\section{Results}
To assess the performance of the selected post-quantum cryptographic algorithms---Kyber512, FrodoKEM, and sntrup761---we conducted 400 iterations of key generation, encryption, and decryption for each algorithm in a controlled test environment. The average execution times for each operation were recorded and are summarized in Tables \ref{tab:execution_time}, \ref{tab:keygen}, and \ref{tab:enc_dec}.

\begin{table}[h]
\centering
\caption{Average Execution Time (in milliseconds)}
\label{tab:execution_time}
\resizebox{\columnwidth}{!}{%
\begin{tabular}{|c|c|c|c|}
\hline
\textbf{Algorithm} & \textbf{Key Generation} & \textbf{Encryption} & \textbf{Decryption} \\ \hline
Kyber512 & 0.0095 ms & 0.0114 ms & 0.0081 ms \\ \hline
FrodoKEM & 0.2301 ms & 0.3181 ms & 0.2989 ms \\ \hline
sntrup761 & 0.1968 ms & 0.0145 ms & 0.0137 ms \\ \hline
\end{tabular}
}
\end{table}

\begin{table}[h]
\centering
\caption{Key Generation Performance}
\label{tab:keygen}
\resizebox{\columnwidth}{!}{%
\begin{tabular}{|c|c|c|}
\hline
\textbf{Algorithm} & \textbf{KeyGen Mean (ms)} & \textbf{KeyGen StdDev} \\ \hline
Kyber512 & 0.0097 & 0.0136 \\ \hline
FrodoKEM & 0.2301 & 0.0411 \\ \hline
sntrup761 & 0.1968 & 0.0446 \\ \hline
\end{tabular}
}
\end{table}

\begin{table}[h]
\centering
\caption{Encryption and Decryption Performance}
\label{tab:enc_dec}
\resizebox{\columnwidth}{!}{%
\begin{tabular}{|c|c|c|c|c|}
\hline
\textbf{Algorithm} & \textbf{Encrypt Mean (ms)} & \textbf{Encrypt StdDev} & \textbf{Decrypt Mean (ms)} & \textbf{Decrypt StdDev} \\ \hline
Kyber512 & 0.0108 & 0.0019 & 0.0079 & 0.0017 \\ \hline
FrodoKEM & 0.3181 & 0.0346 & 0.2989 & 0.0280 \\ \hline
sntrup761 & 0.0145 & 0.0165 & 0.0137 & 0.0048 \\ \hline
\end{tabular}
}
\end{table}
\par
The experimental results show that Kyber512 has the fastest performance across all three operations, making it highly suitable for environments requiring both high security and low latency. FrodoKEM, while offering strong security based on unstructured lattices, has the slowest performance due to its computational complexity. sntrup761 demonstrates a good trade-off between security and performance, especially excelling in encryption and decryption speed, although its key generation is slower compared to Kyber512.
\par
The observed performance metrics confirm the trade-off between security level and computational efficiency in post-quantum algorithms. Kyber512 provides a balanced profile suitable for real-time applications. sntrup761 may be preferred in lightweight environments, while FrodoKEM is better suited for scenarios prioritizing strong resistance over performance.
\par
These results highlight the practical differences between structured and unstructured lattice-based cryptographic approaches and provide a baseline for selecting algorithms in resource-constrained or latency-sensitive applications.

\section{Discussion}
The results presented in this study highlight key differences in performance and efficiency among the three post-quantum cryptographic algorithms---Kyber512, sntrup761, and FrodoKEM. While all three algorithms offer resistance against quantum attacks and are considered secure under current knowledge, they differ significantly in terms of speed, computational overhead, and practical applicability.
\par
Among all evaluated algorithms, Kyber512 consistently demonstrated superior efficiency across all metrics tested. It demonstrated the fastest key generation, encryption, and decryption times, with extremely low standard deviation. These results confirm Kyber's suitability for real-time communication systems and resource-constrained environments. The consistent performance across 400 iterations indicates not only high efficiency but also stability and predictability, which are essential for secure system integration.
\par
sntrup761 also performed well, especially in encryption and decryption, where it achieved nearly comparable speed to Kyber512. However, its key generation time was higher, and standard deviation values showed more variability compared to Kyber. Despite this, sntrup761 remains a strong candidate for practical deployment, especially in lightweight and embedded systems due to its smaller memory footprint and strong security foundation.
\par
FrodoKEM, on the other hand, displayed significantly higher computational costs in all operations. The key generation time was approximately 20x slower than Kyber512, and encryption/decryption times were also much longer. This is expected, as FrodoKEM is based on unstructured lattices, which, while offering potentially stronger security guarantees, inherently result in lower efficiency. The increased resource demand makes FrodoKEM may pose challenges for latency-sensitive applications due to its significantly higher computational overhead. But potentially more appropriate for applications where maximum security is required and performance is less critical.
\par
From a security standpoint, all three algorithms rely on lattice-based constructions, which are believed to be resistant to both classical and quantum adversaries. However, the choice between structured (Kyber, sntrup761) and unstructured (FrodoKEM) lattices poses a trade-off between performance and conservative security assumptions. While Kyber and sntrup761 benefit from optimized structures, FrodoKEM's unstructured approach may offer better resilience against future cryptanalytic breakthroughs, albeit at the cost of speed and efficiency.
\par
Another consideration is implementation complexity and ease of standardization. Kyber512 is already selected for standardization by NIST, making it a frontrunner in the global adoption of PQC. sntrup761 and FrodoKEM, while not selected as primary algorithms, continue to be evaluated in alternate categories and remain valuable candidates for niche use cases or backup algorithms.
\par
In conclusion, while all three algorithms fulfill the primary objective of resisting quantum attacks, Kyber512 offers the most practical balance of security and performance. sntrup761 provides a competitive alternative with moderate resource usage, while FrodoKEM serves as a high-security option where performance trade-offs are acceptable. The choice of algorithm should therefore be guided by the specific requirements of the target application, including processing capabilities, latency tolerance, and security level.
\section{Conclusion}
As quantum computing progresses toward practical implementation, classical cryptographic systems such as RSA and ECC are at risk of becoming obsolete due to their vulnerability to quantum algorithms. In response to this emerging threat, post-quantum cryptographic algorithms like Kyber512, sntrup761, and FrodoKEM have been developed to provide quantum-resistant security.
\par
This study conducted a comprehensive performance evaluation of these three algorithms under consistent conditions. Our results show that Kyber512 offers the best overall performance, with minimal computation time and high consistency, making it an ideal candidate for real-world applications requiring both speed and security. sntrup761 performed similarly well in encryption and decryption operations, though it demonstrated higher variability and slower key generation. FrodoKEM, while significantly slower, provides a more conservative approach using unstructured lattices, which may offer additional security against future threats.
\par
Ultimately, the selection of a post-quantum algorithm depends on the specific requirements of the target system. Systems that prioritize speed and scalability may adopt Kyber512, while applications requiring lightweight operations may benefit from sntrup761. For maximum-security applications where performance can be sacrificed, FrodoKEM may be the appropriate choice.
\par
This comparison provides a foundation for understanding the strengths and weaknesses of each approach and contributes to the broader effort of preparing secure infrastructures for the post-quantum era.

\section{Future Work}
Future work in the area of post-quantum cryptography should focus on several critical aspects:

\begin{itemize}
    
    \item \textbf{Hybrid Integration:} Combining classical and post-quantum algorithms to ensure security during the transition period.
    \item \textbf{Real-world Deployment Testing:} Implementing these algorithms in existing communication protocols (e.g., TLS, VPNs, messaging apps) to test their performance in realistic scenarios.
    \item \textbf{Algorithmic Optimization:} Improving the performance of slow but secure algorithms like FrodoKEM without compromising their security assumptions.
    \item \textbf{Quantum Attack Simulation:} Developing better models and simulations to assess how each algorithm withstands future quantum adversaries beyond current assumptions.
    \item \textbf{Hardware-Level Integration:} Evaluating performance on embedded systems, IoT devices, and low-power processors to ensure applicability across diverse platforms.
    \item \textbf{Standardization and Global Policy:} Supporting global efforts (like NIST PQC standardization) to unify and accelerate the adoption of quantum-safe cryptographic solutions.
\end{itemize}
\par
As quantum threats continue to evolve, ongoing research and proactive implementation strategies will be essential in achieving a secure transition to the post-quantum cryptographic landscape.

\end{document}